\documentclass[%
 reprint,
 amsmath,amssymb,
 aps,
]{revtex4-2}

\usepackage{graphicx}% Include figure files
\usepackage{dcolumn}% Align table columns on decimal point
\usepackage{bm}% bold math
\usepackage{xcolor}
\usepackage{algorithm} 
\usepackage{algpseudocode} 

\begin{document}

%\preprint{APS/123-QED}

\title{Numerical study on the deadline-concerning priority queuing model}% Force line breaks with \\

\author{Hang-Hyun Jo}
\email{h2jo@catholic.ac.kr}
\affiliation{Department of Physics, The Catholic University of Korea, Bucheon 14662, Republic of Korea}

\date{\today}% It is always \today, today,

\begin{abstract}
The Barab\'asi's priority queuing model [A.-L. Barab\'asi, Nature \textbf{435}, 207 (2005)] and its variants have been extensively studied to understand heavy-tailed distributions of the inter-event times and the response times observed in various empirical analyses of human dynamics. In this paper, we focus on the effects of deadlines assigned to the tasks in a queue of fixed size on the response-time distributions. Here, the response time is defined as the time interval between the arrival and the execution of the task. We propose a deadline-concerning priority queuing model, in which as the deadline approaches, the priority is adjusted using the inverse of the remaining time to the deadline. By performing the numerical simulations, we find that the power-law exponent characterizing the response-time distributions is less than $1$ under the deterministic selection protocol while it has the value of $1$ under the nondeterministic selection protocol.
\end{abstract}

\maketitle

\section{Introduction}

Recently, a huge amount of digital data on human activities and social interaction has enabled us to explore human behavior in an unprecedented way~\cite{Lazer2009Computational}. Among many other topics, non-Poissonian or heterogeneous temporal patterns of human individuals have been extensively studied for last few decades in terms of bursty human dynamics~\cite{Karsai2018Bursty}. The bursty temporal patterns have been characterized by using heavy-tailed distributions of inter-event times and response times. Here, the inter-event time denotes the time interval between two consecutive events in the event sequence while the response time denotes the time interval between arrival and execution of, e.g., a task in the queue. Barab\'asi reported in his seminal paper~\cite{Barabasi2005Origin} that both the inter-event times and the response times for an individual email user follow a heavy-tailed distribution such as $P(\tau)\sim \tau^{-\alpha}$ with $\alpha\approx 1$, where $\tau$ denotes either the inter-event time or the response time. Since then, a wide range of values of $\alpha$ have been reported based on analyses of a number of empirical datasets for human social phenomena; for example, the value of $\alpha$ ranges from $0.7$~\cite{Karsai2011Small, Karsai2012Correlated} up to $3$~\cite{Jiang2013Calling} for mobile phone communications. For a summary of such empirical results, one can refer to Tables in Chapter 3 of Ref.~\cite{Karsai2018Bursty} and references therein.

To understand the origin of heavy tails in the distributions of the inter-event times and the response times observed in the email communication dataset, Barab\'asi proposed a priority queuing model with a queue of fixed size~\cite{Barabasi2005Origin}: An individual is assumed to have a list of $L$ tasks, each of which is assigned a priority that is randomly drawn from a certain priority distribution. At each time step, with a probability $p$, the task with the highest priority is selected; otherwise, with a probability $1-p$, one of $L$ tasks is uniformly and randomly selected. The selected task is executed and removed from the list, and a new task with a random priority arrives at the list. The time interval between the arrival and the execution of the task defines the response time. Due to the task selection protocol based on the priority assigned to the tasks, some tasks with low priority have to wait longer for execution than those with high priority, leading to a power-law distribution of response times  with the power-law exponent $\alpha=1$. Note that $\alpha=1$ is obtained only when $p\to 1$. Later, this model was exactly solved for the case with $L=2$~\cite{Vazquez2005Exact, Gabrielli2007Invasion}, as well as for the general $L$ case~\cite{Anteneodo2009Exact}. 

A number of variants of the Barab\'asi's priority queuing model have also been studied to understand the diverse scaling behaviors observed in various empirical analyses~\cite{Vazquez2006Modeling, Cajueiro2008Role, Masuda2009Priority, Min2009Waiting, Oliveira2009Impact, Jo2011Emergence, Mryglod2012Editorial, Zhou2017Model, Karsai2018Bursty, Mukherjee2018Priority, Lee2021Selforganized}. For example, a time-varying priority of the task was considered to understand the editorial review processes of papers published in Physical Review journals~\cite{Jo2012Timevarying}. The priority of the task can vary depending on the context or situation. One common situation in real life is the deadline assigned to the task. The effects of a deadline for the conference on the registration behavior have been empirically studied~\cite{Alfi2007Conference, Alfi2009How, Flandrin2010Empirical} while how such deadlines affect the response-time distribution for the tasks has not yet been fully explored, to our knowledge. In this paper, to understand the effects of deadlines assigned to the tasks in a queue of fixed size on the response-time distribution, we introduce a deadline-concerning priority queuing model and numerically study the model to find the scaling behavior of the response-time distributions.

\section{Model}

\begin{figure}[!t]
\includegraphics[width=\columnwidth]{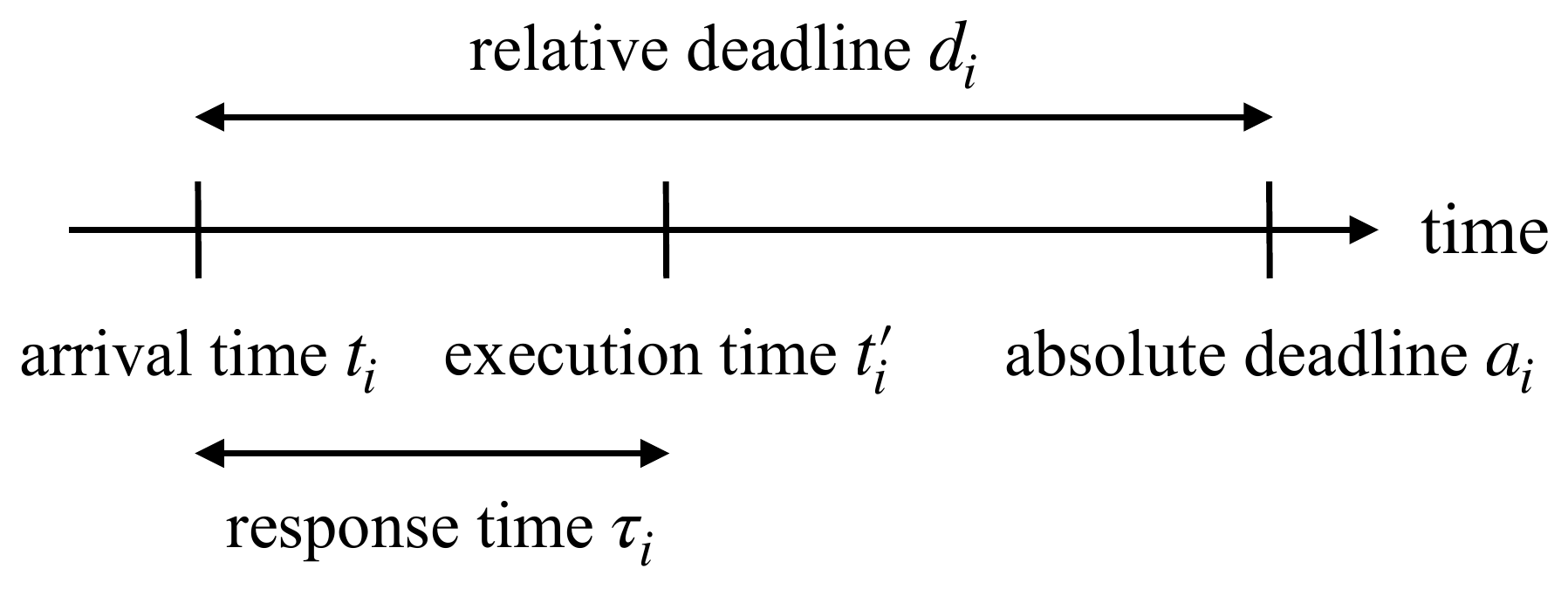}
\caption{Definition of the terms for task $i$ used in the model: Task $i$ with a random priority $x_i$ and a random relative deadline $d_i$ arrives in time $t_i$ and is executed in time $t'_i$. Note that the absolute deadline is given as $a_i= t_i+d_i$ and that the response time is defined as $\tau_i= t'_i-t_i$.}
\label{fig:scheme}
\end{figure}

We study the effects of the deadlines assigned to the tasks on their response times. We consider a model for a queue with a fixed size $L$. Each task $i$ ($i=1,\ldots,L$) in the queue arrives in discrete time $t_i$ with a random priority $x_i$ and a random relative deadline $d_i$, leading to the absolute deadline being given as $a_i=t_i+d_i$ (see Fig.~\ref{fig:scheme}). Note that $d_i$ is a positive integer and that $t_i=0$ for tasks that are initially given. The priority and the relative deadline are randomly drawn from the priority distribution $P(x)$ and the relative deadline distribution $P(d)$, respectively. In the beginning of each time step $t$, all tasks that meet the absolute deadlines, i.e., $\{i|a_i=t\}$, are removed from the queue without execution; then, they are replaced by the same number of newly arrived tasks with random priorities and random relative deadlines. The priority of each task $i$ is adjusted using the information on how much time is left until the absolute deadline for $t_i\le t<a_i$:
\begin{align}
    \hat x_i(t)\equiv \frac{x_i}{a_i-t}.
    \label{eq:adj_prior}
\end{align}
The inverse proportion of the remaining time to the absolute deadline in Eq.~\eqref{eq:adj_prior} implies that as the absolute deadline approaches, the effective priority of the task increases. Then, with probability $p$, the task with the highest adjusted priority is selected among $L$ tasks; otherwise, one of $L$ tasks is uniformly and randomly selected with probability $1-p$. The selected task is executed at the end of the time step $t$, and a new task arrives in the beginning of the time step $t+1$ with a random priority and a random relative deadline. For the executed task $i$, the response time is calculated as $\tau_i= t-t_i$. This procedure is repeated until the time step reaches $T$.

The case with $p=1$ implies a deterministic selection of the task for execution while $p=0$ means a fully random selection of the task. In our work, we will call the case with $0<p<1$ the nondeterministic selection protocol. We remark that the inverse proportion of the remaining time to the absolute deadline in Eq.~\eqref{eq:adj_prior} has been introduced to explain the temporal behavior of conference registrations and fee payments~\cite{Alfi2007Conference, Alfi2009How, Flandrin2010Empirical}. We also note that some tasks can be executed as soon as they arrive, implying $\tau_i=0$. Finally, for the executed task $i$, $\tau_i< d_i$ by definition. 

In our work, we focus on the case with uniform distributions for both $x_i$ and $d_i$ as follows:
\begin{align}
    &P(x)=1\quad \textrm{for}\ x\in[0,1],\\
    &P(d)=\frac{1}{D}\quad \textrm{for}\ d=1,\ldots,D.
\end{align}
Therefore, the maximum response time is set to $D-1$.

\section{Numerical results}

\subsection{Deterministic case with $p=1$}

\begin{figure}[!t]
\includegraphics[width=\columnwidth]{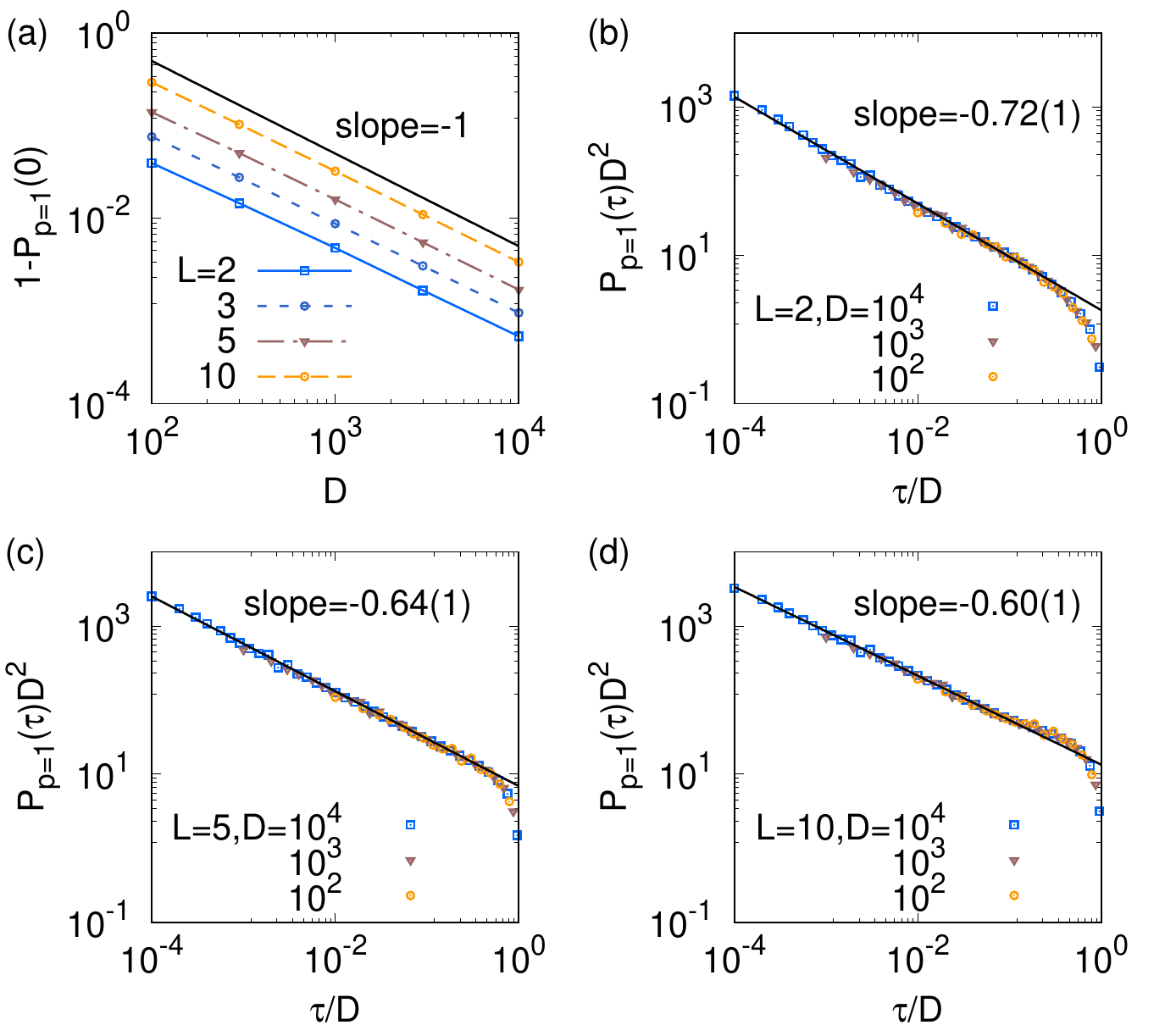}
\caption{Response-time distributions $P_{p=1}(\tau)$ of the model for various combinations of $L$ and $D$ in the deterministic case with $p=1$. In panel (a) we show the fractions of tasks with $\tau=0$ as a function of $D$ for various values of $L$. In panels (b--d) we show that the distributions $P_{p=1}(\tau)$ collapse onto a single curve when $P_{p=1}(\tau)D^2$s are plotted against $\tau/D$ for all values of $L$ considered.}
\label{fig:r_p1}
\end{figure}

We first study the case with $p=1$, i.e., the deterministic case in which the task with the highest adjusted priority is always selected for execution at each time step. We numerically simulate the model up to time $T=10^9$ for various combinations of $L$ and $D$ to obtain the response-time distribution $P_{p=1}(\tau)$, and the result is shown in Fig.~\ref{fig:r_p1}. We find that $\tau=0$ is dominant in all combinations of $L$ and $D$. For each $L$, the fraction of $\tau=0$, i.e., $P_{p=1}(0)$, increases with the increasing $D$, which can be approximated as
\begin{align}
    P_{p=1}(0)\approx 1-\frac{c_L}{D}
    \label{eq:P0}
\end{align}
with an $L$-dependent coefficient $c_L$, as shown in Fig.~\ref{fig:r_p1}(a). Precisely, we get $c_L\propto L^{1.22(8)}$. For a fixed $L$, in the limit of $D\to\infty$, one obtains $P_{p=1}(0)\to 1$. The distributions for the range $\tau>0$ can be characterized in the following scaling form with a normalization constant $C$:
\begin{align}
    P_{p=1}(\tau)=C\tau^{-\alpha}f\left(\frac{\tau}{D}\right).
    \label{eq:Ptau}
\end{align}
Here, $f$ denotes a cutoff function, which can be approximated as $f(u)=\theta(1-u)$ with $\theta$ denoting a Heaviside step function. Naturally, the cutoff is found to be linear in $D$ because by the definition of the model $\tau$ is limited only by $D$. The power-law exponent $\alpha$ characterizing the scaling behavior of the response-time distributions for $D=10^4$ is estimated as $\approx 0.72(1)$ for $L=2$, $\approx 0.64(1)$ for $L=5$, and $\approx 0.60(1)$ for $L=10$, see Fig.~\ref{fig:r_p1}(b--d). In all cases, we find $\alpha$ to be less than one, which has been rarely reported in the literature~\cite{Karsai2018Bursty}: $\alpha\approx 0.7$ was observed for the inter-event time distributions in mobile phone-call communications~\cite{Karsai2011Small, Karsai2012Correlated}. In addition, $\alpha<1$ for the response-time distributions was suggested in a model study considering the cost and the benefit of communication~\cite{Jo2012Optimized}.

The normalization condition of $P_{p=1}(\tau)$ reads
\begin{align}
    \sum_{\tau=0}^{D-1} P_{p=1}(\tau)=1.
    \label{eq:norm}
\end{align}
Plugging Eqs.~\eqref{eq:P0}~and~\eqref{eq:Ptau} into Eq.~\eqref{eq:norm}, one gets for $D\gg 1$
\begin{align}
    C\approx c_L(1-\alpha)D^{\alpha-2},
\end{align}
leading to 
\begin{align}
    P_{p=1}(\tau)\propto D^{-2}\left(\frac{\tau}{D}\right)^{-\alpha}f\left(\frac{\tau}{D}\right)
\end{align}
for $\tau\ge 1$. Therefore, the distributions for different values of $D$ are expected to collapse onto a single curve when $P_{p=1}(\tau)D^2$ is plotted against $\tau/D$, which turns out to be the case, as shown in Fig.~\ref{fig:r_p1}(b--d).

\begin{figure}[!t]
\includegraphics[width=\columnwidth]{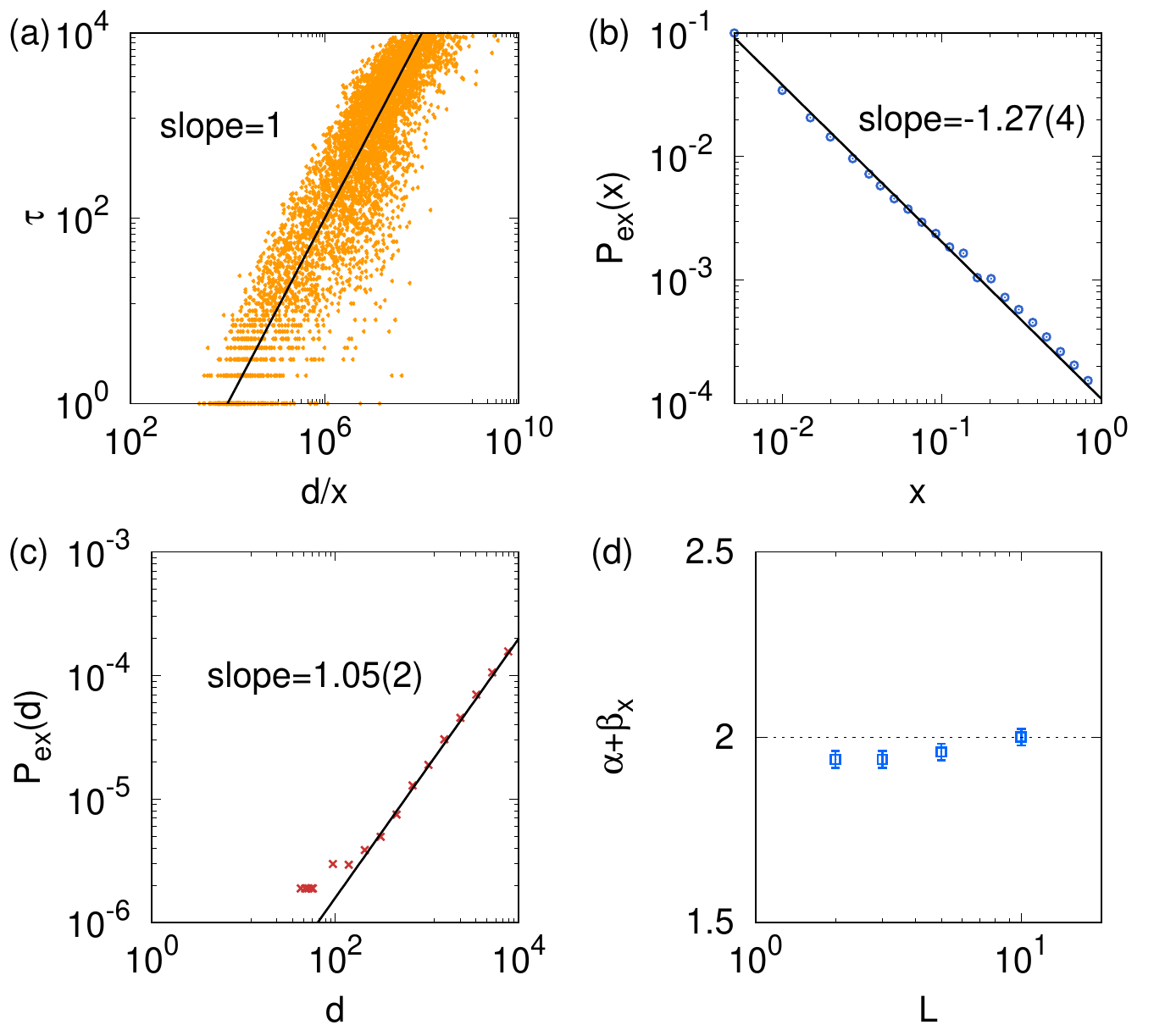}
\caption{(a--c) Statistics for the executed tasks with $\tau\ge 1$ in the deterministic case with $p=1$, where we used $L=2$, $D=10^4$, and $T=10^9$; scatter plot of ($d/x$, $\tau$) for the first five thousands executed tasks (a), the distribution of $x$ for all executed tasks (b), and the distribution of $d$ for all executed tasks (c). (d) The estimated value of $\alpha+\beta_x$ as a function of $L$ for the case with $D=10^4$ and $T=10^9$. The error bars denote the standard deviations.}
\label{fig:xdr}
\end{figure}

For understanding the results, we obtain the statistics for the executed tasks. We expect a task with longer relative deadline and smaller priority to wait longer for execution, implying a larger response time. In Fig.~\ref{fig:xdr}(a), we find a positive (or roughly linear) correlation between $d/x$ and $\tau$, as expected. We also measure the distributions of $x$ and $d$ for executed tasks (Fig.~\ref{fig:xdr}(b,~c)) and find that $P_{\rm ex}(x)\sim x^{-\beta_x}$ with $\beta_x\approx 1.27(4)$ and that $P_{\rm ex}(d)\sim d^{\beta_d}$ with $\beta_d\approx 1.05(2)$. As the relative deadline $d$ affects the response time $\tau$ linearly, we focus on the effect of the priority $x$ on $\tau$. Using the identity relation $P(\tau)d\tau=P_{\rm ex}(x)dx$ together with $P_{\rm ex}(x)\sim x^{-\beta_x}$ and $\tau\propto 1/x$, one gets 
\begin{align}
    P(\tau)\sim \tau^{-(2-\beta_x)},
\end{align} 
implying that
\begin{align}
    \alpha=2-\beta_x.
\end{align}
This relation is confirmed by the estimated values of $\alpha$ and $\beta_x$ from the numerical simulations for the various values of $L$ when $D=10^4$ is used, as shown in Fig.~\ref{fig:xdr}(d).

\subsection{Nondeterministic case with $0<p<1$}

\begin{figure}[!t]
\includegraphics[width=\columnwidth]{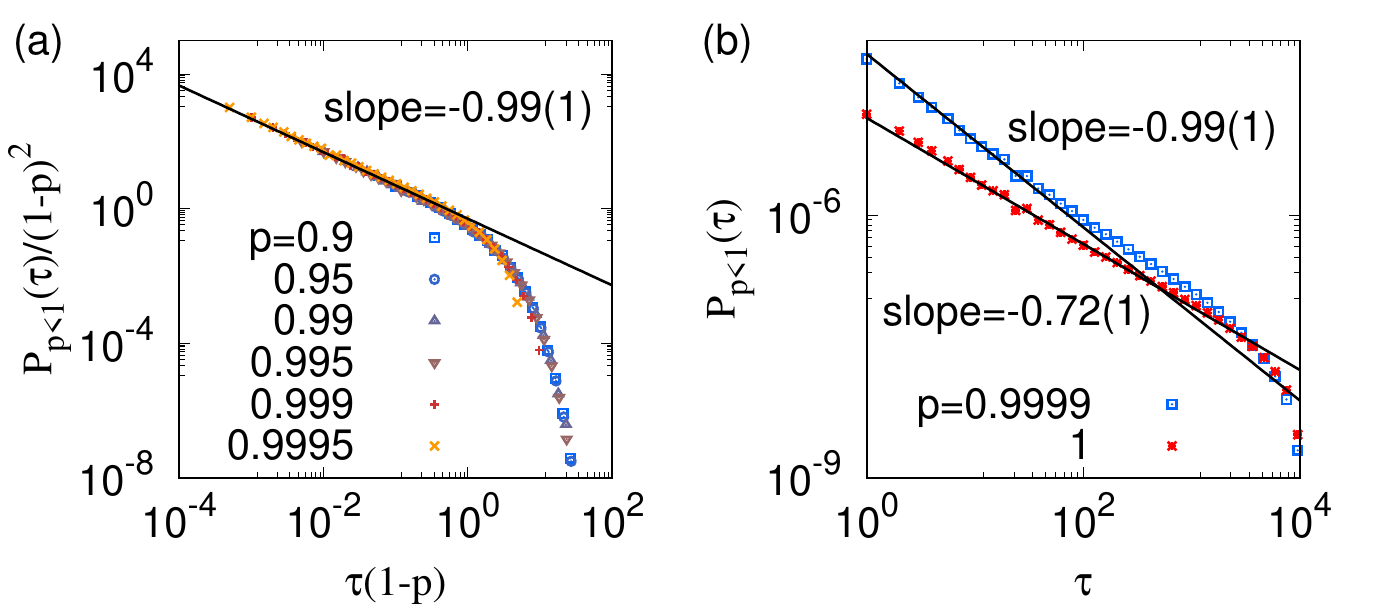}
\caption{Response-time distributions $P_{p<1}(\tau)$ of the model for various values of $p$ close to $1$ when $L=2$, $D=10^4$, and $T=10^9$ are used. (a) The distributions for $p\le 0.9995$ collapse onto a single curve when the values of $P_{p<1}(\tau)/(1-p)^2$ are plotted against $\tau(1-p)$. (b) On the other hand, the distribution for $p\ge 0.9999$ develops a hump for the range of $10^2<\tau<10^3$, which is comparable to $P_{p=1}(\tau)$.}
\label{fig:r_L2}
\end{figure}

Next, we study the effect of stochastic selection of the task for the case with $p<1$. We numerically obtain the response-time distributions for various values of $p$ close to $1$ for the case with $L=2$ and $D=10^4$, and we show the result in Fig.~\ref{fig:r_L2}. In contrast to the deterministic case, a cutoff appears in the distribution due to the finite value of $p$ as long as the order of $(1-p)^{-1}$ is smaller than that of $D$. As shown in Fig.~\ref{fig:r_L2}(a), the response-time distributions for various values of $p\le 0.9995$ can be characterized in the following form:
\begin{align}
    P_{p<1}(\tau)\propto (1-p)^2[\tau(1-p)]^{-\alpha}g[\tau(1-p)],
\end{align}
where $g$ denotes an exponential cutoff function. Here, the power-law exponent $\alpha$ is found to have a value of $\approx 0.99(1)$. On the other hand, in Fig.~\ref{fig:r_L2}(b), we observe that the distribution for $p\ge 0.9999$ develops a hump in the range of $10^2<\tau<10^3$, leading to a heavier tail characterized by a smaller value of $\alpha$. The distribution for the range of $\tau$ for the hump is found to overlap with that for the case with $p=1$. Therefore, we conclude that the nature of exponential cutoff in the distribution is associated with a different scaling behavior: $\alpha\approx 1$ when the cutoff is determined by $(1-p)^{-1}$, while $\alpha<1$ when the cutoff is determined by $D$. Similar behavior is found for other values of $L$ (not shown). Finally, we remark that in the Barab\'asi's priority queuing model, the power-law tail of the response-time distribution appears only for the range of $p<1$~\cite{Barabasi2005Origin, Vazquez2005Exact}, whereas in our model, the scaling behavior is found even at $p=1$.

\section{Conclusion}

In this paper, we have studied the effects of the deadlines assigned to tasks on the response times by focusing on the scaling behavior of the response-time distributions. We numerically find a power-law exponent less than $1$ under the deterministic selection protocol ($p=1$). Such a scaling behavior can be, to some extent, understood by means of the power-law distribution of the priorities of executed tasks with response times larger than $0$. When the nondeterministic selection protocol is applied ($0<p<1$), an exponential cutoff in the distributions develops as it is stronger than the cutoff due to the maximum relative deadline. The power-law exponent characterizing the scaling behavior of the response-time distribution is found to be around $1$. Hence, we conclude that the value of power-law exponent is associated with the nature of a cutoff. The analytical approach to understanding these numerical results is left for a future work.

\begin{acknowledgments}
H.-H.J. was supported by Basic Science Research Program through the National Research Foundation of Korea (NRF) funded by the Ministry of Education (NRF-2018R1D1A1A09081919).
\end{acknowledgments}

%\appendix

%\section{Appendixes}

%\bibliography{h2jo-papers}% Produces the bibliography via BibTeX.

\end{document}